\documentclass[12pt,preprint]{aastex}

\newcommand{\msun}{\mbox{${\rm M}_\odot$}}

\newcommand{\rsun}{\mbox{${\rm R}_\odot$}}

\newcommand{\kms}{\mbox{${\rm km~s}^{-1}$}}

\def\apgt{\ {\raise-.5ex\hbox{$\buildrel>\over\sim$}}\ }
\def\aplt{\ {\raise-.5ex\hbox{$\buildrel<\over\sim$}}\ }
\def\lt{\ {\raise-.5ex\hbox{$\buildrel>$}}\ }
\def\gt{\ {\raise-.5ex\hbox{$\buildrel<$}}\ }

\newfont{\Giga}{cmssbx10 scaled 5200}
\newfont{\giga}{cmssbx10 scaled 4300}
\newfont{\Mega}{cmssbx10 scaled 3200}
\newfont{\mega}{cmssbx10 scaled 2500}
\newfont{\Kilo}{cmssbx10 scaled 2000}
\newfont{\kilo}{cmssbx10 scaled 1600}
\newfont{\Deca}{cmssbx10 scaled 1450}
\newfont{\deca}{cmssbx10 scaled 1200}
\newfont{\Dezi}{cmssbx10 scaled 1100}
\newfont{\dezi}{cmssbx10 scaled 1050}
\newfont{\iGiga}{cmssi10 scaled 6200}
\newfont{\igiga}{cmssi10 scaled 4300}
\newfont{\iMega}{cmssi10 scaled 3200}
\newfont{\imega}{cmssi10 scaled 2500}
\newfont{\iKilo}{cmssi10 scaled 2000}
\newfont{\ikilo}{cmssi10 scaled 1500}
\newfont{\mathGiga}{cmsy10 scaled 6200}
\newfont{\mathgiga}{cmsy10 scaled 4300}
\newfont{\mathMega}{cmsy10 scaled 3200}
\newfont{\mathmega}{cmsy10 scaled 2500}
\newfont{\mathKilo}{cmsy10 scaled 2000}
\newfont{\mathkilo}{cmsy10 scaled 1500}
\newfont{\mathDeca}{cmsy10 scaled 1450}
\newfont{\mathdeca}{cmsy10 scaled 1200}

\righthead{Formation of planets in triple systems}
\lefthead{S.\ F.\ Portegies Zwart \& S. L. W. McMillan}

\begin{document}
\title{Planets in triple star systems---the case of HD188753}

\author{Simon F.\ Portegies Zwart\altaffilmark{1,2}
\and Stephen L.\ W.\ McMillan\altaffilmark{3}}

\altaffiltext{1}{Astronomical Institute ``Anton Pannekoek,''
University of Amsterdam, Kruislaan 403, 1098 SH Amsterdam, NL}

\altaffiltext{2}{Department of Computer Science,  University of Amsterdam,
Kruislaan 403, 1098 SH Amsterdam, NL}

\altaffiltext{3}{Department of Physics,Drexel University, 
Philadelphia, PA 19104, USA steve@physics.drexel.edu}

\date{Received 2005 August 1; 
      in original form 1687 October 3.6; Accepted xxxx xxx xx.}

\label{firstpage}

\begin{abstract}

We consider the formation of the recently discovered ``hot Jupiter''
planet orbiting the primary component of the triple star system
HD188753.  Although the current outer orbit of the triple is too tight
for a Jupiter-like planet to have formed and migrated to its current
location, the binary may have been much wider in the past.  We assume
here that the planetary system formed in an open star cluster, the
dynamical evolution of which subsequently led to changes in the
system's orbital parameters and binary configuration.  We calculate
cross sections for various scenarios that could have led to the
multiple system currently observed, and conclude that component A of
HD188753 with its planet were most likely formed in isolation to be
swapped in a triple star system by a dynamical encounter in an open
star cluster. We estimate that within 500\,pc of the Sun there are
about 1200 planetary systems which, like Hd188753, have orbital
parameters unfavorable for forming planets but still having a planet,
making it quite possible that the HD188753 system was indeed formed by
a dynamical encounter in an open star cluster.
\end{abstract}

\keywords{methods: N-body simulations -- 
          planets and satellites: individual HD188753 --
	  planetary systems: formation --
	 } 
	  
\section{Introduction}

The recently discovered ``hot Jupiter'' orbiting the main sequence
star HD188753A with period $P_A = 3.3481\pm0.0009$\,days has
significantly complicated our view of planet formation (Konacki
2005)\nocite{2005Natur.436..230K}.  The principal problem is the
presence of the binary companion HD188753B, whose proximity severely
constrains the planet formation process in the HD188753 system.

HD188753 was discovered and recognized as an interesting object---a
stable hierarchical triple system---in the late 1970s (Griffin
1977)\nocite{1977Obs....97...15G}.  The primary HD188753A has mass
$m_A = 1.06\pm0.07$ {\msun}.  The secondary HD188753B is itself a
binary system, consisting of early (4.5\,Gyr old) main sequence stars
of total mass $m_B = m_{B1} + m_{B2} = 0.96 + 0.67 \pm 0.07$\,\msun\,
(Konacki 2005)\nocite{2005Natur.436..230K}.  The inner (B) binary has
period $P_B = 156.0\pm0.1$\,days, semi-major axis $a_B \simeq 0.67$
AU, and eccentricity $e_B = 0.1\pm0.03$ (Konacki
2005)\nocite{2005Natur.436..230K}.  The outer (AB) binary has period
$P_{AB} \simeq 25.7$\,yr, semi-major axis $a_{AB}$ = 12.1 AU, and
eccentricity $e_{AB} \simeq 0.50$.

Jupiter-mass planets form in stellar and binary systems beyond the
``snow line,'' the distance from the parent star where the
protostellar disk is just cool enough for water ice to form Hayashi
(1981, see also Sasselov \& Lecar,
2000)\nocite{1981PThPS..70...35H,2000ApJ...528..995S}.  For solar-mass
stars, the snow line lies at a distance of roughly 3AU; in the
particular case of interest here, it lies at $r_{\rm snow} \simeq
2.7$\,AU from HD188753A (Konacki 2005)\nocite{2005Natur.436..230K}.
Hot Jupiters subsequently sink inward due to interactions with the
disk to a distance of $\sim 10-20$ {\rsun} from the parent star.

Truncation of the disk can prevent the formation of giant planets, and
the formation of a gas-giant planet orbiting HD188753A in its current
configuration would have been significantly hampered by the
perturbation due to HD188753B.  The disk surrounding the A component
should have been truncated at a radius of about 1.4\,AU, substantially
less than the minimum distance ($\sim 2.7$\,AU) at which a gas giant
could form Hayashi (1981)\nocite{1981PThPS..70...35H}\footnote{The
exact location of the snow-line and its actual relevance are discussed
more extensively by Kornet, {R{\'o}{\.z}yczka} \& Stepinski
(2004)\nocite{2004A&A...417..151K}.}.  Thus the classical formation
scenario for the planet fails, raising the question of how HD188753A's
hot Jupiter came into being.

Rather than exploring exotic new regimes in planet formation, this
paper focuses on the more mundane possibility that the hot Jupiter
orbiting HD188753A did in fact form beyond the snow line, in
accordance with the standard theory of the formation of gas-giant
planets.  We propose that the present orbit came about due to a
dynamical encounter in which the binary HD188753B became a companion
of HD188753A {\em after} the planet had already formed and reached its
current orbit close to its parent star.  We present a plausible
scenario whereby this could have occurred and argue that this is the
most promising and least extreme explanation of the HD188753 system.

The constraints on this study are simple.  The planetary system
(HD188753A) and binary (HD188753B) must have survived the interaction
that brought them together, and the snow line in the post-encounter
system must lie outside the present disk truncation radius,
prohibiting {\em in situ} formation of the giant gas planet. We
calculate cross sections for such encounters and discuss the
corresponding formation rates in typical open star clusters.

\section{Formation Scenarios}

Given the difficulty in forming HD188753A in place, we investigate two
alternative formation mechanisms for the HD188753 system:
\begin{itemize}
    \item[$I$)] The planetary system HD188753A formed as an isolated
	      object and entered its current orbit around the binary
	      HD188753B following an exchange interaction.\footnote{An
	      exchange is one possible outcome of a three-body
	      single-star--binary scattering, in which when the
	      incoming star displaces either the primary or the
	      secondary of the binary (see Hut \& Bahcall 1983).}  In
	      this case, HD188753B must originally have had a
	      companion star of unknown mass, which was ejected as a
	      result of the dynamical encounter.  There are no
	      constraints on the formation of the hot-Jupiter planet,
	      as there was no companion to perturb it.
    \item[$II$)] The planetary system HD188753A formed as a companion to
	      another star of unknown mass, but in a sufficiently
	      wide orbit to allow for the formation of the planet.  An
	      encounter with the binary HD188753B subsequently led to
	      an exchange interaction, ejecting the unknown star and
	      placing the binary in orbit around HD188753A.
\end{itemize}

A third, but less likely, possibility is that an existing triple
system was born wide enough to allow the formation of a planet around
the primary component.  Subsequent preservation interactions harden
the outer orbit of the triple by one or more fly-bys, until the
current tight orbit was reached.  This scenario can we writen as: (AB
+ C $\rightarrow$ AB + C).  Our simulations indicate that the cross
sections for this type of encounter are comparable to the other cross
sections presented in Fig.\,\ref{Fig:cross_section} up to an orbital
period of about 10\,yr.  The preservation cross section, however,
continues to rise to a maximum of $\sigma \simeq 7 \times 10^6$AU$^2$
near the hard-soft boundary at an orbital period of about 40\,yr,
after which it drop sharply.  This process could therefore have
comparable weight to the other two, except that it requires a
complicated multiple system with a planet to begin with, which
ultimately may render this channel unimportant

We adopt the parameters described above for the planetary system in
HD188753A and the binary companion HD188753B.  Both the A and B
components are assumed to remain largely unaffected by the dynamical
encounter responsible for the present system---that is, we assume that
the interaction did not result in an encounter close enough to perturb
either binary component.  This simplifying assumption allows us to
model the various interactions as three-body encounters, rather than
considering in detail the potentially very complex 5-body dynamics of
the entire system.  We could of course relax these criteria, but we
regard it as unlikely that such refinements will materially affect our
conclusions.  Since binaries and multiple systems appear to be the
norm in open star clusters (Kouwenhoven et al
2005)\nocite{2005A&A...430..137K}, the assumption that component A or
B formed in orbit around star C does not represent a significant
limitation on our discussion.  On the basis of a limited number of 4-
and 5-body scattering calculations, we conclude that, with the
effective radii adopted below, our 3-body scatterings do not differ
materially from their more detailed counterparts.

We compute cross sections for the relevant exchange encounters by
means of 3-body scattering experiments conducted using the {\tt
scatter3} and {\tt sigma3} programs in the {\tt Starlab} software
environment (McMillan \& Hut, 1996; Portegies Zwart et al
2001)\nocite{1996ApJ...467..348M,2001MNRAS.321..199P}.  Throughout, we
adopt notation in which object A is HD188753A and its planet, object B
is the binary HD188753B, and C is the unknown third star involved in
the dynamical interactions responsible for the HD188753 system we now
see.  The masses of the three objects are denoted $m_A$, $m_B$, and
$m_C$, and their (effective) radii are $r_A$, $r_B$, and $r_C$.  The
semi-major axis and eccentricity of the AB binary are denoted $a_{AB}$
and $e_{AB}$, with similar notation for the BC or AC binaries should
they arise.

We vary the mass $m_C$ of the third component in the initial triple
between 0.1 and 1\,\msun, and assign it an appropriate zero-age
main-sequence radius.  The effective radii of components A and B are
taken to be $r_A$ = 0.05\,AU and $r_B$ = 0.67\,AU, ensuring minimal
perturbation of these components during the encounters.  For our
purposes, a ``successful'' exchange is defined as one in which
\begin{enumerate}
\item the initial conditions are stable and allow ``normal'' formation
      of a hot Jupiter,
\item objects never approach one another within the sum of their
      effective radii, and
\item the final system is a stable hierarchical system similar to
      HD188753, in which giant planet formation appears forbidden by
      standard planet-formation theories.
\end{enumerate}
These criteria are described in more detail in the following
subsection.  We note in passing that the choice of effective radius
$r_B$ is not simply a matter of our not wishing to perturb the B
system (whose initial conditions we do not know).  Our detailed
scattering experiments indicate that a close encounter between A or C
and the tightly bound binary B can release sufficient energy from B to
suppress the exchange process of interest here.  However, we find that
our choice of $r_B$ in the 3-body approximation is adequate to prevent
this from occurring.

\subsection{Insertion of the planetary system into an existing triple}

In the first series of experiments we consider the possibility that
HD188753A and its planet exchanged into an existing triple system:
A + BC $\rightarrow$ AB + C.  The incomer (A) in our scattering
calculations is HD188753A (including its planet in a 3.3\,day orbit).
The target is a triple system consisting of an inner binary star (B)
and an outer component C of unknown mass $m_C$.

We vary the initial orbital semi-major axis $a_{BC}$ of the outer
binary between 100 and $10^6$\,{\rsun} and the relative encounter
velocity at infinity between 1\,km/s, appropriate for an encounter in
an open cluster, and 10\,km/s, to mimic a more massive parent cluster.
The initial outer binary eccentricity $e_{BC}$ is drawn from a thermal
distribution [$p(e) = 2e$] between $e_{BC}=0$ and $e_{BC}=1$, subject
to additional stability constraints as described below.  The remaining
variables---the line of the ascending node, the eccentric anomaly, the
orbital inclination angle and the moment of periastron passage---are
chosen randomly, as described in Hut \& Bahcall
(1983)\nocite{1983ApJ...268..319H}.

In this scenario there are no strong constraints on the initial
conditions.  The incoming star is isolated, so the presence of a
planet places no restriction on its properties, and the target
hierarchical triple system need only be dynamically stable.  A
coplanar prograde triple system (with particles 1 and 2 forming the
inner orbit, and particle 3 the outer component) is taken to be stable
if the periastron separation of the outer orbit satisfies
\begin{equation}
  p_{out} \apgt 2.8 a_{in} 
	  \left[ (1 + q_{out}) {1+e_{out} 
	    \over (1-e_{out})^{1/2}}) 
	  \right]^{2/5}.
\label{Eq:stable}
\end{equation}
(Mardling \& Aarseth 1999)\nocite{1999dsbs.conf..385M}.  Here,
$q_{out} = m_3/(m_1+m_2)$ is the mass ratio of the outer binary,
$a_{in}$ is the semi-major axis of the inner binary, and $e_{out}$ is
the outer orbital eccentricity.  Configurations with nonzero
inclination between the inner and outer orbits are expected to be more
stable than the planar prograde case (Mardling \& Aarseth
2001)\nocite{2001MNRAS.321..398M}, so a binary radius based on Eq.\
(\ref{Eq:stable}) overestimates the effective size of the B binary
system and hence underestimates the cross sections computed here.

Some of the resulting triple systems have highly inclined outer
orbits, and may be unstable against Kozai resonances, causing the
components of the multiple system ultimately to merge (Ford, Kozinsky,
\& Rasio 2000; Lee \& Peale 2003). However, we find that only a small
fraction ($\aplt 10\%$) of our final systems could be lost by this
mechanism (see also Heggie \& Rasio 1996).\nocite{1996MNRAS.282.1064H}

Condition (3) above places two constraints on the orbital elements of
the final system.  First, the maximum size of a circumstellar disk
after the encounter can be approximated as
\begin{equation}
  R_{\rm disk} \simeq 0.733 R_{\rm L} \left( 1-e \right)^{1.20} 
                                      \left( 1+q \right)^{-0.07}.
\end{equation}
Pichardo, Sparke \& Aguilar (2005)\nocite{2005MNRAS.359..521P}.  Here
$q = m_B/m_A$, $R_L$ is the Roche radius of component $A$, and the
eccentricity dependence results from numerical scattering
calculations.  A binary born with a maximum disk radius smaller than
the snow line ($R_{\rm disk} < R_{\rm snow}$) cannot form gas-giant
planets, and our final configurations therefore must satisfy this
condition.  The second constraint is imposed by the requirement that
the final triple system also be stable, according to
Eq.\,\ref{Eq:stable}.  This criterion is least accurate for small mass
ratios and inclined outer orbits, but we find that our cross sections
are quite insensitive to the details of the stability criterion.

The thick curves in Figure \ref{Fig:cross_section} show the results of
our cross section calculations for scenario I, with a range of initial
orbital periods and companion masses.  The cross sections peak at
initial binary periods of around 15--20 yr, largely independent of the
mass of companion C.  The cutoff at small periods stems from the
stability criterion on the initial BC binary; systems with periods
greater than $\sim10^5$ years have too little energy for exchange
reactions to form the observed AB system.

\subsection{Insertion of a close binary into a wide binary system}

In the second scenario we consider the possibility that star A and its
planetary system formed in a wide binary system with a companion star
C with orbital semi-major axis $a_{AC}$ large enough (by the criterion
just described) for planet formation not to be significantly affected
by C's presence.  In this view, the binary $B$ exchanged into this
binary to form the observed triple system: AC + B $\rightarrow$ AB +
C.  As before, we use scattering experiments to determine the cross
section for such an exchange encounter, now subject to the constraints
(i) that the initial AC binary has $R_{\rm disk} > R_{\rm snow}$ and
(ii) that the final AB binary have $R_{\rm disk} < R_{\rm snow}$ and
satisfy the triple stability criterion.  For definiteness we adopt a
radius of 10\,AU as being sufficiently large for a giant planet to
form and subsequently migrate into a hot Jupiter orbit (Nelson 2000;
Alibert et al. 2005)\nocite{2000ApJ...537L..65N,2005A&A...434..343A}.

The cross sections for scenario II are shown as the thin curves in
Figure\,\ref{Fig:cross_section}.  Apart from a slight shift to shorter
periods and a reduction of a factor of roughly of two in overall
probability, the results are qualitatively similar to those for
scenario I.

To assess the importance of condition (3), we performed additional
simulations while ignoring this restriction. These less restricted
encounters result in cross sections which are larger by a factor of
$\sim 4$ for initial periods of up to $\sim 10$\,yr, rising to $\sim
10$ times larger for binaries with periods exceeding $\sim 100$\,yr.
The peak cross section without condition (3) is about $\sigma \simeq 7
\times 10^6$AU$^2$ at an orbital period of about 40\,years.

We find that, since the vast majority of encounters are hard,
subsequent interactions with other cluster members are unlikely to
widen (or ionize) the final system again to the point where it no
longer satisfies condition (3)---that is, where it no longer appears
``unusual'' in having a planet.  Effective loss of ''successful''
(HD188753-like) systems is achieved only by coalescence during a
resonant encounter, and those are relatively rare.  During such an
encounter, however, it is quite possible that all stars and the planet
coalesce (Fregeau et al 2004).\nocite{2004MNRAS.352....1F}

\section{Discussion and Conclusions}

We can combine the cross sections presented in Figure
\ref{Fig:cross_section} with the binary period distribution reported
by \nocite{1991A&A...248..485D} to estimate how often encounters of
the sort described here have occurred within $\sim 100$\,pc of the
sun, and hence how many ``HD188753-like'' objects one might expect in
the solar neighborhood.  The period-averaged cross section for
scenario I is $\sigma_I \simeq 1.8 \times 10^5\,$AU$^2$, $3.4 \times
10^5\,$AU$^2$ and $4.8 \times 10^5\,$AU$^2$ for $m_C = 0.1\,\msun$,
0.5 and 1.0\,\msun, respectively.  For scenario $II$ we find
$\sigma_{II} \simeq 1.0 \times 10^5\,$AU$^2$, $2.0\times 10^5\,$AU$^2$
and $3.2 \times 10^5\,$AU$^2$ for $m_C = 0.1$\,\msun, 0.5 and
1.0\,\msun.

The overall rate for encounters of type I in a cluster of $N_\star$
stars with mean number density $n$ may be expressed as
\begin{equation}
	\Gamma_I = f_I N_\star n \sigma_I v\,.
\end{equation}
Here $f_I \equiv f_A f_{BC}$, where $f_A$ is the fraction of hot
Jupiter systems like HD188753A and $f_{BC}$ is the fraction of stable
triples containing a binary similar to B.  We can similarly calculate
$\Gamma_{II}$ by substituting $\sigma_{II}$ in $\sigma_I$ and $f_{II}
\equiv f_{AC}f_B$ in $f_I$, where $f_B$ the fraction of short-period
binaries and $f_{AC}$ the fraction of wide binaries in which one
component has a hot Jupiter.

The values of $f_A$, $f_{B}$, $f_{AC}$ and $f_{BC}$ are not trivial to
assess and therefore we retain them as parameters in the equations.
The fraction of single cluster stars hosting a hot Jupiter $f_A$ is
not well known, but we can at least estimate the other fractional
parameters.  The fraction of short period binaries is $f_{B} \sim
0.25$ (Duquennoy \& Mayor 1991; Pourbaix et
al. 2005)\nocite{1991A&A...248..485D,2005yCat.5122....0P}, the
fraction of binaries with a hot-Jupiter in orbit around the primary is
$f_{AC} \sim 0.1f_A$, and the fraction of triples is $f_{BC} \apgt
0.1$ (Tokovinin 1999)\nocite{1999yCat..41240075T}.  We then estimate
$f_I \aplt 0.1f_A$, $f_{II} \sim 0.025f_A$.

For simplicity we assume that the secondary masses in a binary system
are distributed with equal probability between 0.1\,\msun\, and the
mass of the primary, in which case the average cross sections become
$\sigma_I \simeq 3.3\times 10^5$AU$^2$ and $\sigma_{II} \simeq
2.0\times 10^5$AU$^2$.  The total number of HD188753-like systems in
the solar neighborhood then is $N_{HD} \sim \Gamma N_c t$, where
$\Gamma \equiv \Gamma_I+\Gamma_{II}$ is the total rate for scenarios I
and II, $N_c$ is the total number of open star clusters within the
volume of interest, and $T$ is a characteristic cluster lifetime.
After substitution and scaling to parameters typical of open clusters,
we obtain
\begin{equation}
	N_{\rm HD} \simeq 0.23 \left(n \over [pc^{-3}]\right)
	                     \left(f_I\sigma_I + f_{II}\sigma_{II}
	                     \over [10^5AU^2]\right)
			     \left(v \over [km/s]\right) 
			     \left(N_\star \over [100]\right)
			     \left(N_c \over [1]\right)
			     \left(T \over [Gyr]\right).
\label{Eq::Number}\end{equation}
Here $n \sim 1$\,pc$^{-3}$ is the stellar density of an open cluster
and $v \sim 1$\,\kms\, is the cluster's velocity dispersion.  Note
that all clusters born over the last $\sim 8$\,Gyr contribute, as the
observed system is dynamically stable and the most massive star
requires at least this time to leave the main-sequence.  

The open cluster catalog of Kharchenko et al.~ (2005)
\nocite{2005A&A...438.1163K} gives ages, distances, core radii and
observed membership for 520 open clusters, $\sim 90\%$ of which lie
within 2\,kpc of the Sun.  Here we consider only the 79 clusters lying
within 500\,pc listed in the Kharchenko et al., (2005) catalog. For
each of these clusters we compute the stellar density from the core
radius and the number of stars, and the velocity dispersion (adopting
a mean mass of 0.5\,\msun) from the number of cluster members and the
core radius from the catalog.  The average stellar density the cluster
in this sample is $n \simeq 6.7$\,stars pc$^{-3}$; the mean number of
cataloged stars per cluster is $N_\star \simeq 35$.  If we then adopt
$f_I\sigma_I + f_{II}\sigma_{II} \simeq 5.3\times 10^4$AU$^2$ and
sum the contribution for each of the clusters in the Kharchenko et
al. catalog we arrive at about one planetary system with
characteristics similar to HD\,188753 within 500\,pc of the Sun.

However, the Kharchenko et al. catalog is rather incomplete.  They
mainly used data from the ASCC-2.5 full-sky survey, which is complete
down to $V=11.5$\,mag (Kharchenko et al. 2005).  At a distance of
100--500\,pc this magnitude corresponds to that of a 0.8--1.5\,\msun\,
main-sequence star. For a Kroupa (2002) \nocite{2002ASPC..285...86K}
initial mass function, 87--95\% of the stars are less massive than
this, ignoring stellar remnants, and would be unaccounted for in the
Kharchenko et al. catalog. We correct for this by increasing the
number of stars in the cluster catalog by an order of magnitude to
make up for unobserved stars.  Summing the modified contributions for
all clusters results in a total of $\sim 170$ planetary systems like
HD188753.

An additional correction should be made for the star clusters which
are not in the catalog.  Kharchenko et al (2005) estimate that there
are about 1700 open star clusters within a kpc of the sun (see also
Lynga 1987; Loktin et al. 1997)\nocite{Lynga1987,1997BaltA...6..316L},
whereas the Kharchenko et al (2005) catalog only lists 234 within that
volume.  Applying both corrections---for unobserved cluster members
and incompleteness in the catalog---we estimate that, within 500\,pc
of the Sun, there are about 1200 planetary systems which, like
HD188753, have orbital parameters apparently unfavorable for forming
giant planets, but which nevertheless contain such a planet.

\acknowledgments 

This paper was written during the MODEST-5C summer school in Amsterdam
which was supported by the Dutch Organization for Scientific Research
(NWO), the Dutch Advanced School for Astronomy (NOVA), and the Royal
Netherlands Academy of Arts and Sciences (KNAW).  We also acknowledge
the support of NASA's Astrophysics Theory Program under grant
NNG04GL50G, and thank the other instructors at the school: Douglas
Heggie, Piet Hut, James Lombardi, Bill Paxton, and Peter Teuben.

\bigskip
{\bf Note added in proof:} A recent submission of Eric Pfahl to ApJ Letters
(astro-ph/0509490) performs similar calculations and arrives at
comparable conclusions to those presented this paper.

\clearpage

\begin{figure}
\plotone{f1.eps}
\caption[]{Cross sections (in units of $10^5$\,AU$^2$) for encounters
of type $I$ (top set of thick curves) and type $II$ (lower thin
curves), as functions of initial binary period.  The line styles
identify the experiments performed for various choices of the
secondary mass $m_C$ in the initial system, which was varied from $m_C
= 0.1$\,{\msun} (solid curves), 0.5\,{\msun} (dashed curves) and
1.0\,{\msun} (dotted curves).  The velocity at infinity was fixed at 1
km/s.  The error bars at the left indicate the 2-sigma uncertainties
in the graphs (upper: scenario I, lower: scenario II).
\label{Fig:cross_section}
}
\end{figure}

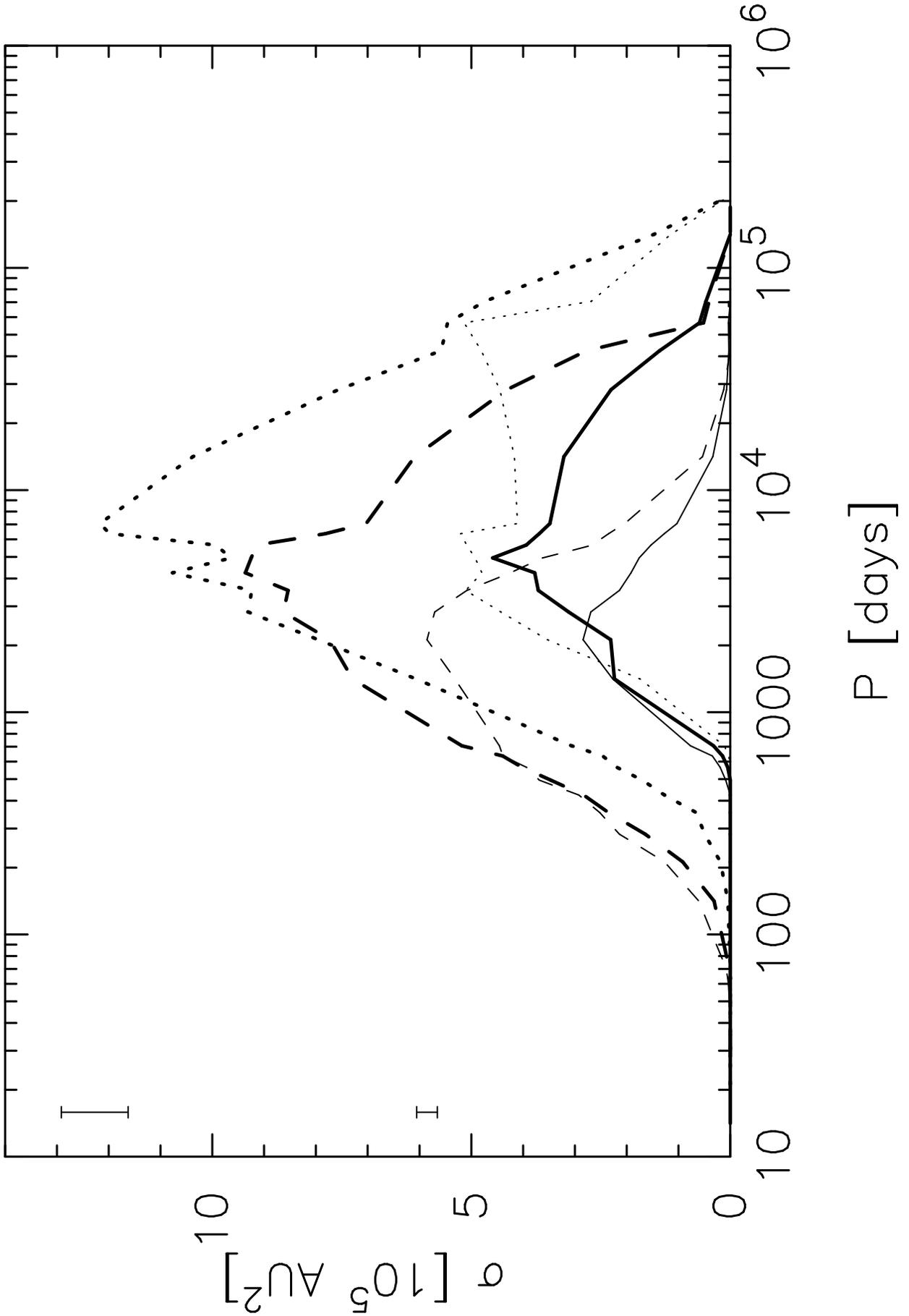
\end{document}